\documentclass{aa}    

\usepackage{epsfig}
\usepackage{url}
\usepackage{txfonts}
\usepackage[breaklinks=true]{hyperref}  
\usepackage{breakurl}
\usepackage{graphicx}
\usepackage{fixltx2e}

\hypersetup{
  colorlinks=true,
  citecolor=blue,
  linkcolor=blue,
  urlcolor=blue}

\usepackage{natbib,twoopt}
\bibpunct{(}{)}{;}{a}{}{,}   
\makeatletter
 \newcommandtwoopt{\citeads}[3][][]{\href{http://adsabs.harvard.edu/abs/#3}%
   {\def\hyper@linkstart##1##2{}%
    \let\hyper@linkend\@empty\citealp[#1][#2]{#3}}}   
 \newcommandtwoopt{\citepads}[3][][]{\href{http://adsabs.harvard.edu/abs/#3}%
   {\def\hyper@linkstart##1##2{}%
    \let\hyper@linkend\@empty\citep[#1][#2]{#3}}}      
 \newcommandtwoopt{\citetads}[3][][]{\href{http://adsabs.harvard.edu/abs/#3}%
   {\def\hyper@linkstart##1##2{}%
    \let\hyper@linkend\@empty\citet[#1][#2]{#3}}}   
 \newcommandtwoopt{\citeyearads}[3][][]%
   {\href{http://adsabs.harvard.edu/abs/#3}%
   {\def\hyper@linkstart##1##2{}%
    \let\hyper@linkend\@empty\citeyear[#1][#2]{#3}}} 
\makeatother

\begin{document}  

\title{Solar extreme ultraviolet variability of the quiet Sun}

\author{F. Shakeri\inst{1}
  \and L. Teriaca\inst{1}  
  \and S. K. Solanki\inst{1,2}}

\institute{Max-Planck-Institut f\"ur Sonnensystemforschung, 37077, G\"ottingen, Germany, 
  \and School of Space Research Kyung Hee University, Yongin, Gyeonggi-Do, 446-701, Korea}

\abstract {The last solar minimum has been unusually quiet compared to the previous minima (since space-based radiometric measurements are available). The Sun's magnetic flux was substantially lower during this minimum. Some studies also show that the total solar irradiance during the minimum after cycle 23 may have dropped below the values known from the two minima prior to that. For chromospheric and coronal radiation, the situation is less clear-cut. The Sun's 10.7\,cm flux shows a decrease of $\sim4\%$ during the solar minimum in 2008 compared to the previous minimum, but \ion{Ca}{II} K does not. Here we consider additional wavelengths in the extreme ultraviolet (EUV), specifically transitions of \ion{He}{I} at 584.3\,\AA\ and \ion{O}{V} at 629.7\,\AA , of which the CDS spectrometer aboard SOHO has been taking regular scans along the solar central meridian since 1996. We analysed this unique dataset to verify if and how the radiance distribution undergoes measurable variations between cycle minima. To achieve this aim we determined the radiance distribution of quiet areas around the Sun centre. Concentrating on the last two solar minima, we found out that there is very little variation in the radiance distribution of the chromospheric spectral line \ion{He}{I} between these minima. The same analysis shows a modest, although significant, 4\% variation in the radiance distribution of the transition region spectral line \ion{O}{V}. These results are comparable to those obtained by earlier studies employing other spectral features, and they confirm that chromospheric indices display a small variation, whereas in the TR a more significant reduction of the brighter features is visible.} 

\maketitle

\section{Introduction}
The minimum in solar activity between cycles 23 and 24 was different in many aspects from other recent solar minima. It lasted longer, with nearly twice as many spotless days as during the previous eight minima, and was particularly deep and inactive (\citeads{2011GeoRL..38.6701S}; \citeads{2011SoPh..272..337T}). The Sun's interplanetary field in the recent minimum was at approximately half of the value it had during solar minimum 21 and was at its lowest level since the beginning of the measurements in 1963 (\citeads{2009JGRA..11411104L}). This result is consistent with a halving of the Sun's open flux (\citeads{2010A&A...509A.100V}). 

A particularly interesting question is how the Sun's radiative output reacted to the particularly low level of activity. Of the four main composites of total solar irradiance (TSI), ACRIM (\citeads{2003GeoRL..30.1199W}; \citeads{2009GeoRL..36.5701S}), PMOD (\citeads{2006SSRv..125...53F}; \citeads{2009A&A...501L..27F}), RMIB (\citeads{2008SoPh..247..203M}), and TIM (\citeads{2011GeoRL..38.1706K}), two (PMOD and the ACRIM) display a decrease in the TSI value during the minimum between cycles 23 and 24 compared to the previous minimum (\citeads{2010ASPC..428...63W}; \citeads{2012SGeo...33..453F}). This decrease is independently obtained by models of solar irradiance (\citeads{2014A&A...570A..85Y}), which assume that the evolution of the surface magnetic field drives solar irradiance variability (\citeads{2013ARA&A..51..311S}). Thus for the TSI, which is formed mainly in the photosphere (\citeads{1998A&A...329..747S}), the evidence suggests a significant difference between the two minima.

The situation is less clear-cut for radiation formed in the upper atmosphere. The solar radio microwave flux at 10.7\,cm is commonly regarded as an activity indicator of the upper chromosphere and lower corona. The 10.7\,cm radio flux has two main sources: thermal bremsstrahlung emission and gyroradiation; therefore, the enhancements in the temperature and the magnetic field of the Sun, as results from emerging sunspots, increase the 10.7\,cm flux, thereby producing a high correlation between the sunspot number and solar radio flux. However, in line with anomalous trends in some other solar activity indices during the last solar cycle, this well-pronounced correlation seems to have changed during solar cycle 23 \citepads{2011SoPh..272..337T}, especially close to the solar minimum in 2008 \citepads{2010ASPC..428..325S}. In addition, two different studies by \citetads{2011CoSka..41..113F} and \citetads{2010ApJ...711L..84T} show a very slight decrease (within 4\%) in the amplitude of the 10.7\,cm flux over the last solar minimum compared to the minimum in 1996. This is tiny compared with the changes displayed by other activity indicators between the two activity minima.

Other solar activity indices, such as extreme ultraviolet (EUV) flux (\citeads{2010GeoRL..3716103S}; \citeads{2010ASPC..428...63W}) and the Mg II index \citep{weber2013}, also exhibit a similar decreasing trend over solar minimum 23. The measurements from the SOHO Solar EUV Monitor (SEM) (\citeads{1998SoPh..177..161J}), for instance, reveal a 15\% decrease with 6\% uncertainty in the irradiance of the spectral band from 26 to 34 nm over minimum 23 with respect to the minimum 22 (\citeads{2010ASPC..428...73D}). 

Particularly in the equatorial region, the solar surface is dominated by two main components: the quiet Sun and the active regions. The magnetic fields emerging from the solar photosphere through the boundaries between super granulation cells continually alter the upper solar atmosphere, resulting in a non-homogeneous EUV brightening (\citeads{2004GMS...141..341F}). Although there are debates about various heating mechanisms of the chromosphere, a common view is that the main origin of the radiation from the chromosphere in the quiet Sun, especially in the internetwork regions, is the dissipation of the acoustic waves, generated in the photosphere as a result of convective motions of the plasma material at the top of the convective zone (\citeads{1991SoPh..134...15R}; \citeads{1997LNP...489..159C}; \citeads{2007ASPC..368...93W}; \citeads{2012A&A...540A.130S}). However, the co-spatiality between chromospheric and photospheric network patterns, although not one-to-one (\citeads{1989ApJ...337..964S}; \citeads{2009IAUS..259..185L}), suggest that a magnetic heating mechanism is at the root of the network formation in the higher atmosphere (\citeads{2012RSPTA.370.3129R}). 

The aim of this paper is to find out whether there is a change in the radiance of the upper chromosphere and transition region of the quiet Sun network and internetwork between the activity minima following cycles 22 and 23. We present the results of a statistical study of the radiance variability of the chromospheric \ion{He}{I} 584.3\,\AA\ ($\sim$20000\,K) and transition region \ion{O}{V} 629.7\,\AA\ ($\sim$250000\,K) spectral lines in the quiet Sun close to disc centre, over a period of 15 years from 1996 to 2011, with the main focus on comparing the last two solar minima. We used observations from the Coronal Diagnostic Spectrometer (CDS) aboard the Solar and Heliospheric Observatory (SOHO), which provides daily observations of the Sun in various wavelengths and has a long enough record covering the last two solar minima. To have a homogenous and consistent data set, all the measurements, during which an active region or bright feature was passing through the field of view of the instrument have been excluded from the final data set. This assures, to the maximum possible extent, that the results are based purely on observations of the quiet Sun. 

\section{Observational data}
The CDS instrument \citepads{1995SoPh..162..233H} aboard SOHO has been observing the Sun since 1996. The instrument consists of two spectrometers, the normal incidence spectrometer (NIS) and the grazing incidence spectrometer (GIS). The gratings in the NIS spectrometer disperse the incident radiation onto the NIS detector in two bands, covering the wavelength range from 310\,\AA\ to 380\,\AA\ (NIS-1) and from 517\,\AA\ to 633\,\AA\ (NIS-2). The spectral pixel size of the NIS ranges from 0.070\,\AA\ at 310\,\AA\ to 0.118\,\AA\ at 630\,\AA . The spectral resolution was initially 0.35\,\AA\ for NIS-1 and 0.5\,\AA\ for NIS-2 in terms of full width at half-maximum (FWHM) \citepads{2010A&A...518A..49D}. In June 1998 the contact with SOHO was lost due to operational errors. After recovery in September 1998, the spectral profiles were broadened and asymmetric, resulting in a reduction in the spectral resolution of NIS (see \citeads{2001ApOpt..40.6292P}). 

\begin{figure}
  \resizebox{\hsize}{!}{\includegraphics{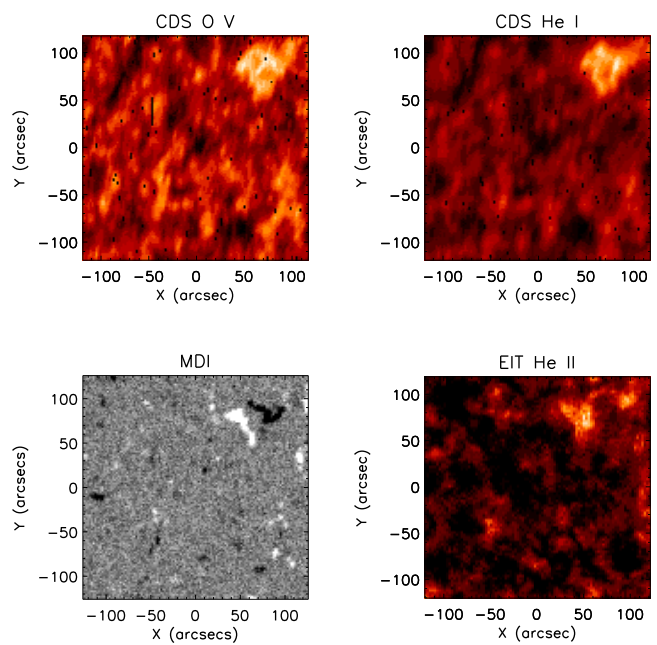}}
  \caption{Appearance of a bright structure in the field of view of the CDS. The MDI magnetograms and EIT images were cropped in order to cover the same area of the solar surface as visible in the CDS images. Images acquired on 1996 July 27}
  \label{4BP}
\end{figure}

In this study, we used NIS daily synoptic observations (the so-called study program SYNOP\_F). These observations cover the central meridian of the Sun from pole to pole. Each run of SYNOP\_F consists of a series of nine raster scans, each covering an area of $240\arcsec \times240 \arcsec$ of the solar disc. Observations were taken employing the $2\arcsec \times240\arcsec$ slit (slit number 4) and by scanning along the solar \textit{x}-direction in 120 separate exposures. The scale of the raster scans is 2\arcsec\ in solar \textit{x}-direction and 1.68\arcsec\ in solar \textit{y}-direction and each raster contains $120\times143$ spectra. In order to concentrate on the quiet Sun area close to disc centre, we only analysed the central rasters of each set of 9, with solar \textit{y} pointings between $-120\arcsec$ and $+120 \arcsec$. The complete data set we analysed covers a period of time from 1996 April 1 to 2011 May 21 and consists of more than 4000 raster scans of daily measurements. As the main goal of this investigation is to look for any statistical variations in the radiance distribution of the quiet Sun during the last two solar minima, we chose a time span of 1 year during each minimum and did a comparative study over various statistical parameters (as will be described in section ~\ref{analyses}). The 1-year periods were chosen to be from 1996 April 1 to 1997 April 1 (minimum 22) and from 2008 July 1 to 2009 July 1 (minimum 23). 

As mentioned above, a selection process was carried out to exclude the rasters scans containing any structure deemed to be unusually bright for the quiet Sun (e.g. due to decaying ARs). The selection was done in two ways: firstly, by visual inspection of radiance images of CDS in the \ion{He}{I} and \ion{O}{V} lines and by referring to the He II 304\,\AA\ images of the Extreme-Ultraviolet Imaging Telescope (EIT) \citepads{1995SoPh..162..291D} and the Michelson Doppler Imager (MDI) \citepads{1995SoPh..162..129S} level 1.8 full-disc magnetograms of the synoptic 96-minute series, and secondly, by analysing the goodness of the lognormal fits to the radiance distributions. This method will be explained in more details in section~\ref{lognormal}.  Fig.~\ref{4BP} illustrates four images acquired by the three instruments on 1996 July 27. The two upper CDS images show a small active region (AR) in the upper right corner. The same feature is also visible in the MDI and EIT images. Simultaneous comparison of the four images makes it possible to identify the raster scans containing bright structures with high confidence. 

\begin{figure*}
\centering
{\includegraphics[width=17cm]{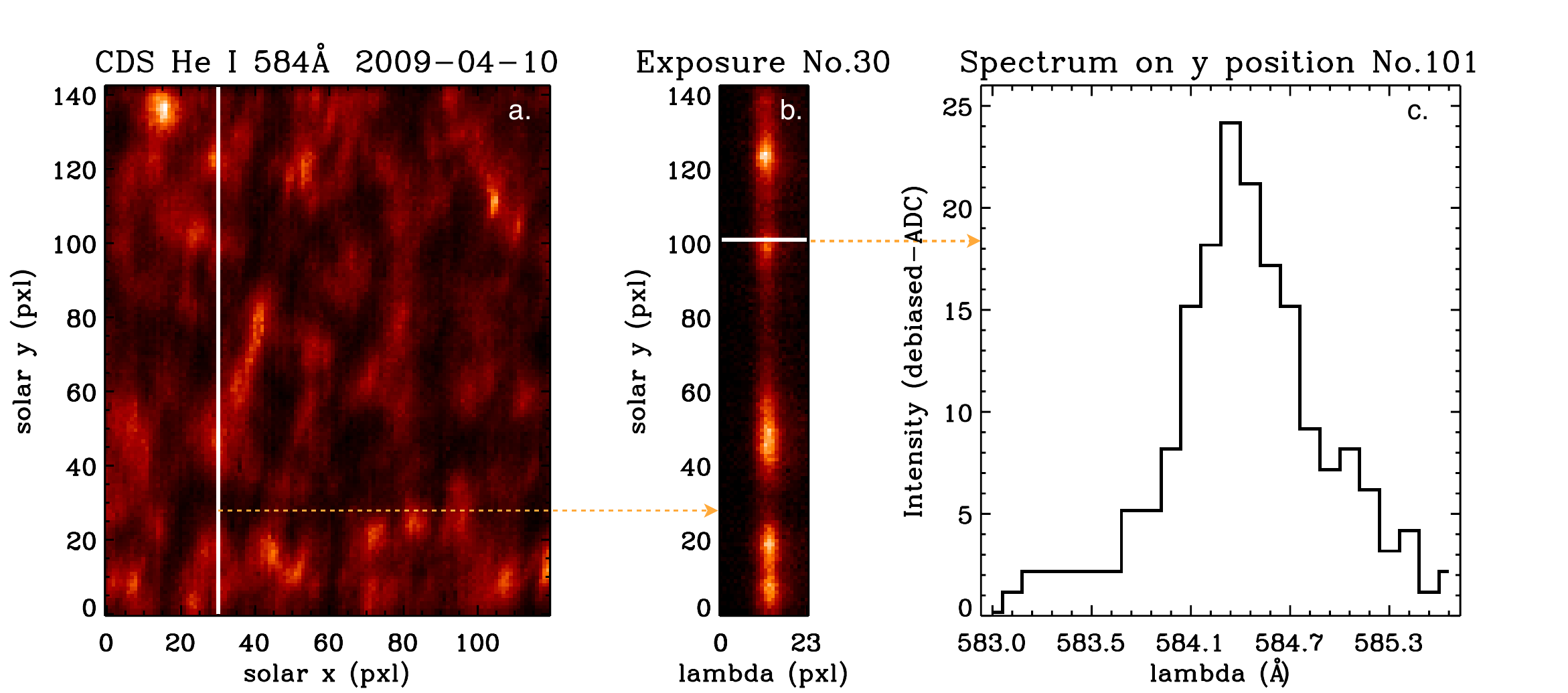}}
  \caption{Panel \textit{a} is an intensity image in \ion{He}{I} spectral line. Each vertical column in the image is corresponding to the average of the spectra over the wavelength dimension for a given \textit{x} position (i.e. exposure). Panel \textit{b} is a given exposure, corresponding to the white line in panel \textit{a}, consisting of 143 spectra along the \textit{y} direction. Panel \textit{c} is a spectrum recorded along the white line in Panel \textit{b}.}
  \label{adc_sp}
\end{figure*}

\subsection{Data reduction}\label{integration}
The CDS raster scans were reduced using a set of \textit{SolarSoft} routines. We applied VDS\_CALIB for de-biasing, flat-fielding and conversion of instrument counts into physical units. We notice here that, in the case of spatially resolved observation, the measured physical quantity are spectral radiance (in units of mW\,sr$^{-1}$\,m$^{-2}$\,nm$^{-1}$) and, after integration over wavelength, radiance (mW\,sr$^{-1}$\,m$^{-2}$). However, it is customary (albeit formally incorrect) to use the word intensity in the place of radiance. We will also make use of the term intensity, always meaning radiance. The cosmic ray hits were removed by applying the standard routine CDS\_NEW\_SPIKE. The CDS$\backslash$NIS data are stored in a cubic structure with the dimensions of  \textit{x}, \textit{y} and \textit{$\lambda$}, corresponding to solar \textit{x} and \textit{y} directions and the wavelength, meaning that each spatial "pixel" of a given raster scan contains a spectrum (see Fig.~\ref{adc_sp}). We analysed individual spectra and excluded the ones which are highly damaged due to cosmic ray hits. On average, in a typical raster scan, approximately 3\% of the spectra were flagged out. NIS spectra also suffer from geometrical distortions such as rotation and tilt of the spectra, resulting from the misalignments between the detector and the grating, and between the grating and the slit, respectively. These effects were corrected by applying the NIS\_ROTATE routine. After applying the instrumental standard calibration and correction routines, the solar radiance at each spatial pixel was determined by least-squares-fitting of a single Gaussian function with constant background to a pre-loss spectrum, i.e. any spectrum recorded prior to June 24, 1998, and by fitting a broadened Gaussian line profile to each post-loss spectrum\footnote{See CDS Software Note Nr.53: \url{http://solar.bnsc.rl.ac.uk/swnotes/cds_swnote_53.pdf}} and finally by extracting the fitting parameters. The final output of the whole process is a set of daily intensity images, with the intensity at each pixel corresponding to the area under the Gaussian after subtracting the background. 

\begin{figure}
  \resizebox{\hsize}{!}{\includegraphics{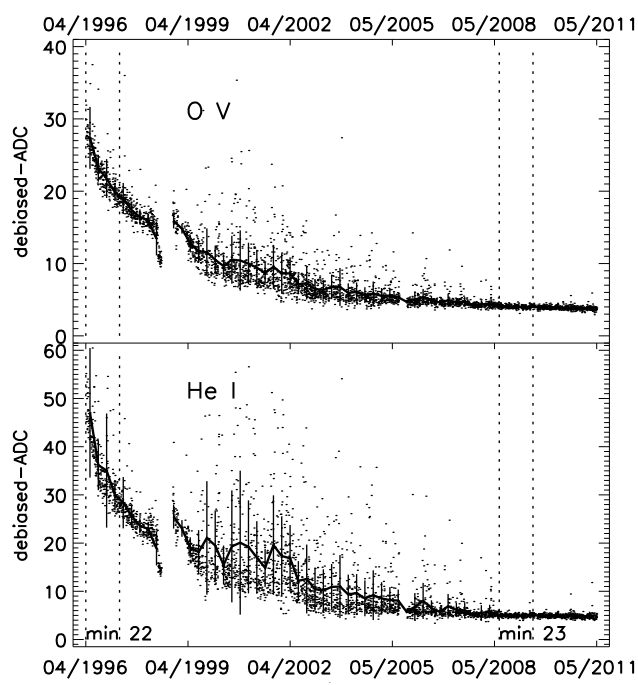}}
  \caption{Illustration of long-term drop in CDS$\backslash$NIS responsivity. The CCD readout bias has been removed from the data right after reading them in. No other calibration has been applied to the data. The solid lines and their associated error bars represent the mean and the standard deviation of the average intensity values over 90-day time intervals. The vertical dashed-lines indicate the two one-year time spans which we used to compare the two solar activity minima.}
  \label{adc}
\end{figure}
\begin{figure}
  \resizebox{\hsize}{!}{\includegraphics{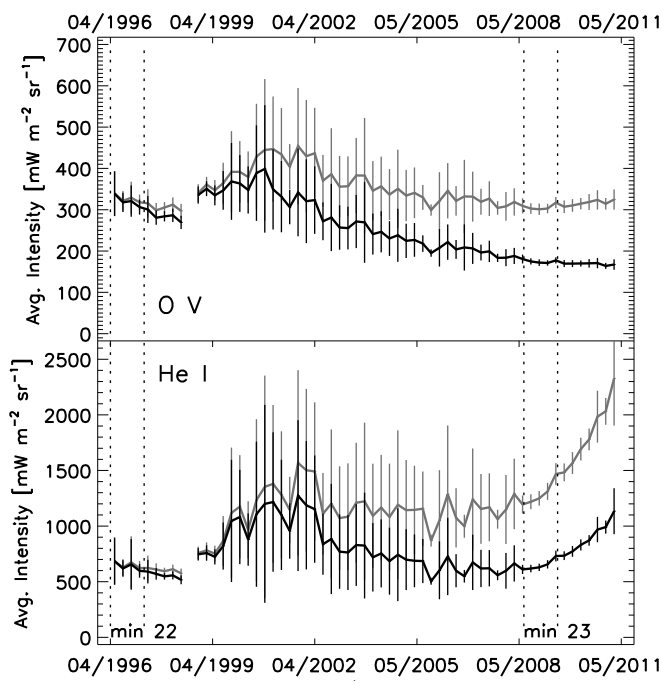}}
  \caption{Long-term variation in daily average intensities of the \ion{He}{I} and \ion{O}{V} spectral lines. The solid lines and their associated error bars represent the mean and the standard deviation of the average intensity values over 90-day time intervals before (grey) and after (black) applying the Del Zanna et al. correction factors. The vertical dashed-lines indicate the two one-year time spans chosen to cover the solar activity minima. }
  \label{avInt}
\end{figure}

\subsection{CDS point-spread function and data binning}\label{psf}
By intercalibration of SUMER detector A and CDS$\backslash$NIS, \citet{Pauluhn..99} pointed out that CDS has a broad PSF corresponding to \textit{FWHM\textsubscript{x}\ = 6\arcsec} and \textit{FWHM\textsubscript{y} = 8\arcsec}, which is almost three times larger than the pixel size of the NIS detector in the \textit{x} direction and over four times larger in the \textit{y} direction. This study was based on observations prior to SOHO loss of control and the post-recovery data are believed to have an even larger point-spread function. Given the broad PSF, binning the intensity images sounds logical, since it increases the signal to noise ratio without reducing the spatial resolution. 
We binned NIS data with six different binsize values: from binsize equal to 2 (binning over groups of $2\times2$ pixels) up to binsize equal to 7 (binning over groups of $7\times7$ pixels). The down-sampling (binning) of the data was carried out after applying instrument calibration corrections and prior to spectra fitting. In this way, we prepared seven data sets corresponding to the different binsize values. The binsize 1 in the plots in this paper, means that the images have not been down-sampled and are in the original format.

\section{Analyses and results}\label{analyses}
\subsection{Daily average intensities}
The CDS/NIS suffers from long-term degradation and decrease in responsivity. To verify the drop in the instrument responsivity, we read in the raw data, removed CCD read-out bias and before applying any further calibration, averaged them over spatial and spectral dimensions. Panel \textit{a} in Fig.~\ref{adc_sp} displays an intensity image in the \ion{He}{I} spectral line in debiased\_ADC unit, i.e. after debiasing and prior to applying calibrations. Each \textit{x} position on the image corresponds to the average of an exposure over the \textit{$\lambda$} dimension. Panel \textit{b} shows a given exposure corresponding to \textit{x} position number 30 of the intensity image (white vertical line in Panel \textit{a}). Panel \textit{c} depicts a spectrum recorded along the white line in panel \textit{b} at \textit{y} position number 101, meaning that the value given to the pixel coordinate [30,101] in panel \textit{a} is the average of the spectrum in panel \textit{c}. The daily average intensities with the unit of "debiased-ADC" were calculated by averaging over \textit{x}, \textit{y}, and \textit{$\lambda$} dimensions. These values are illustrated in Fig.~\ref{adc}, revealing a drastic drop in the responsivity of the instrument over the years. To compensate for this systematic drop, a set of calibration and correction routines has been produced and updated by the CDS team. 

After applying standard calibrations to the data, the daily average intensities in physical units were calculated by fitting the spectra (as explained in section~\ref{integration}) and averaging the intensity in the spectral line over the entire intensity images. The grey line in Fig.~\ref{avInt} illustrates the long-term variations in average intensities in the \ion{He}{I} and \ion{O}{V} spectral lines after applying standard calibrations. It represents the mean intensity values over 90-day time intervals and the vertical bars represent its corresponding standard deviation, respectively. \citetads{2010A&A...518A..49D} suggested that the instrument standard calibration routines overcorrect the long-term drop in the responsivity of the instrument, resulting in a steady increase in the calculated radiance. By analysing the NIS daily synoptic observations from 1996 to 2009, they introduced a new set of correction factors to fix this issue. The black solid line and its corresponding error bars in Fig.~\ref{avInt} display the 90-day averaged intensities after applying the \citetads{2010A&A...518A..49D} correction factors.

The average \ion{He}{I} intensity values calculated by using only the standard corrections (grey line in Fig.~\ref{avInt}), demonstrate an increase starting approximately from 2005. Such a trend is not seen in \ion{O}{V}. Applying Del Zanna et al. correction factors (black lines) moderates the issue to a great extent for \ion{He}{I} but then the \ion{O}{V} intensity values seem to be overcorrected for the period after 2008.

The correction factors introduced by Del Zanna et al. have been determined based on the assumption that the variations in the radiances of the chromospheric and transition region spectral lines are mainly due to the existence of the ARs and that the mean radiances of the spectral lines in the quiet Sun at disc centre essentially does not respond to the solar activity cycle. Although it is expected the solar radiance to be of the same order of magnitude at different solar minima, this assumption essentially precludes any investigation such as the present one, which aims at detecting even small variations in the quiet Sun's radiance, as long as they are significant. The small differences in radiance properties of the quiet Sun between two solar minima can be revealed only if the solar activity parameter under investigation is adequately independent of instrumental degradation. The average intensity values do not seem to satisfy this pre-requisite. The alternative is to identify and study parameters of the intensity distribution which are not affected by instrumental degradation. This is the approach we decided to follow.

\begin{figure}
 \centering
   \resizebox{\hsize}{!}{ \includegraphics{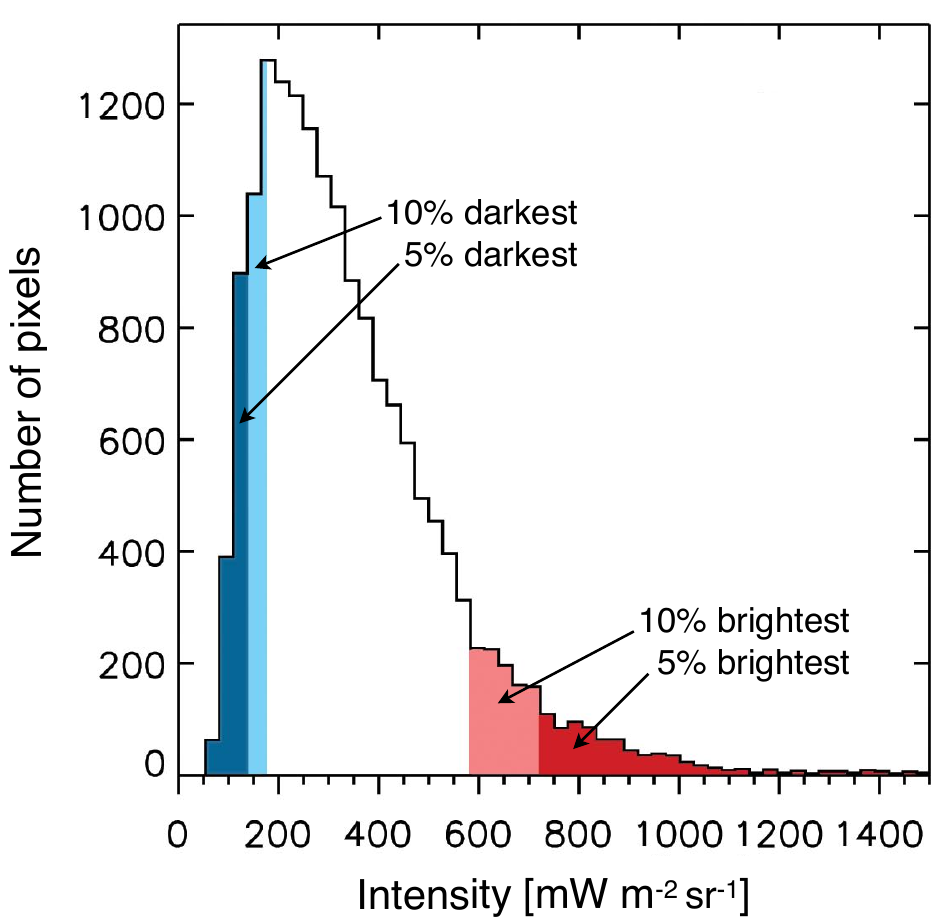}}
  \caption{Histogram gives the number of pixels displaying intensity within a certain range on 1996 June 6. The 5\% pixels with the lowest intensity are shaded dark blue. These plus the lightly shaded blue pixels represent the 10\% darkest pixels. Similarly the more strongly shaded red pixels represent the 5\% brightest pixels. The 10\% brightest pixels are the sum of the pixels shaded in dark and light red. }
  \label{crh}
\end{figure}

\begin{figure*}
 \sidecaption
  {\includegraphics[width=12cm]{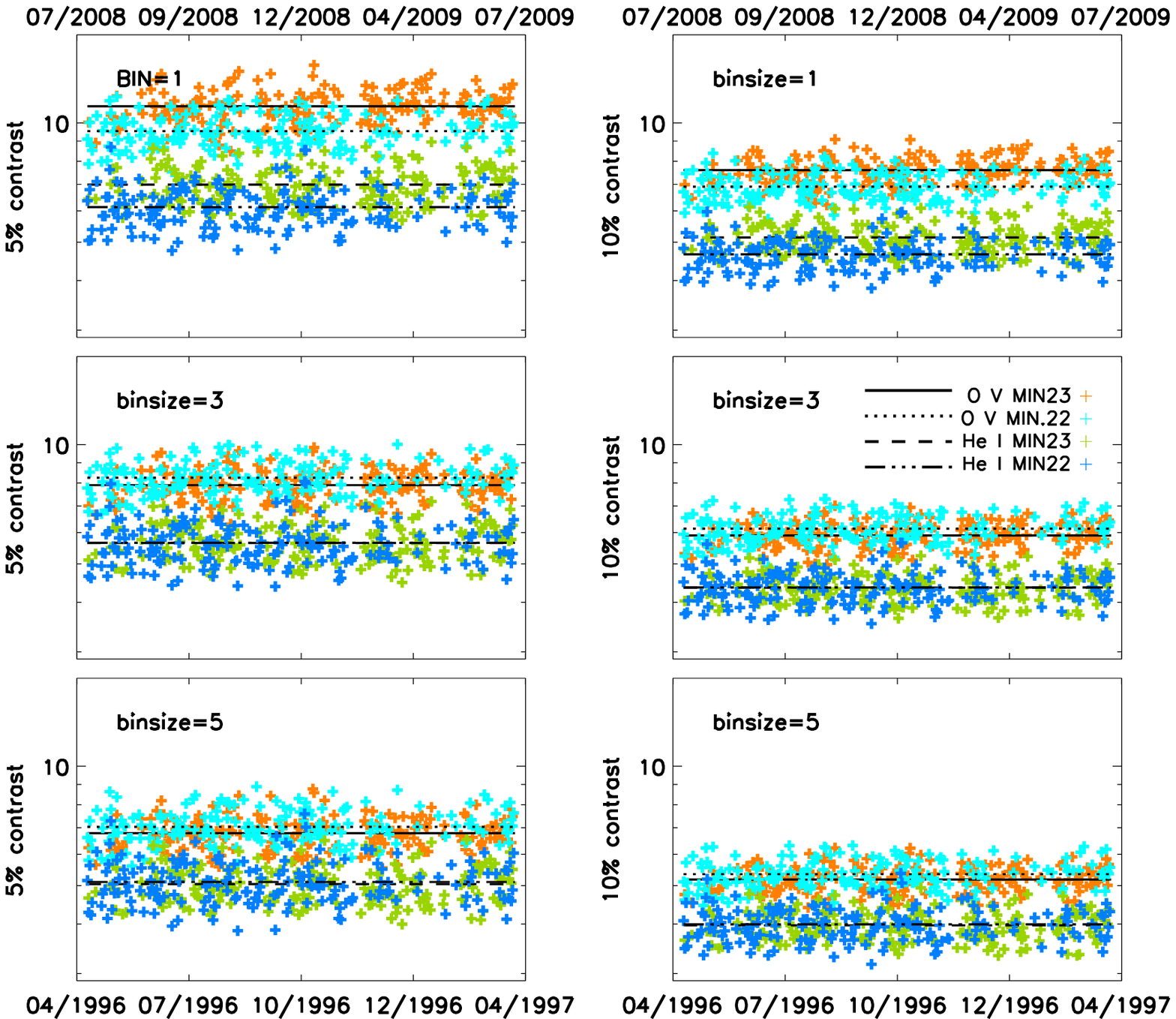}}
  \caption{5\% (left column) and 10\% (right column) contrast ratios (CR) vs. time. See the main text for the definition and Fig~\ref{crh} for an illustration. The one-year periods during minimum 22 and minimum 23 are given by the bottom and top axis, respectively. Each panel represents results for a different binsize (labelled in the panel). The horizontal lines show the average value of each CR. Binsize 1 means that the intensity images are not binned.}
  \label{1year_cnt}
\end{figure*}

\subsection{Contrast ratios}\label{contrast}
In order to minimize the effects of the calibration factors in the analysis, we introduce two ratios, the 5 and 10 percent contrast ratios (called 5\% and 10\% CR in the following), defined as the ratios between the average intensities of the brightest 5\% and 10\% pixels within an intensity map and the average intensities of the darkest 5\% and 10\% pixels, respectively. Fig~\ref{crh} displays a histogram corresponding to a given intensity map. The brightest and darkest pixels are highlighted in red and blue, respectively. The advantage of defining these CRs is that the correction factors which have already been applied to the data will be cancelled out by dividing. From now on we confine our analysis to the solar minima. Fig.~\ref{1year_cnt} illustrates the CRs for three different binsizes (see section~\ref{psf}). In each panel of the figure the four colours distinguish between CRs of the two spectral lines at each minimum. The \textit{x} axes of the bottom and top of each column of the figure represent the date during solar minima 22 and 23, respectively. Excluded from the final plotted dataset, are days on which the CDS images sampled not purely quiet Sun (see Fig.~\ref{4BP}) and/or because substantial parts of the image were missing.

Obviously the 5\% CRs are always larger than the 10\% CRs. This is not surprising, since averaging over a larger number of pixels (10\% vs. 5\%) at the bright and dark extremes of the intensity distribution results in closer mean values and consequently a closer-to-unity ratio. In addition, the CRs decrease as the binsize increases. This is also expected due to the fact that the intensity images tend to reduce towards grey images as more pixels are binned together. It is also evident that for all binsizes \ion{O}{V} shows higher CRs than \ion{He}{I}. This difference starts from approximately 50\% higher CR values for binsize 1, and drops as binsize increases. The possible reasons for these differences will be discussed in detail in the discussion section. What is more in focus here is the differences in CRs of the same spectral line over the two solar minima, e.g. differences between the solid and dotted horizontal lines (for \ion{O}{V}) or between the dashed and dash-dotted horizontal lines (for \ion{He}{I}). For a binsize of 1, this difference is quite large for both spectral lines. However with increasing binsizes, it diminishes in \ion{He}{I}, but persists in \ion{O}{V}. 

\begin{figure}
 \centering
   \resizebox{\hsize}{!}{ \includegraphics[trim=0.1cm 0cm 2.35cm 0cm, clip=true]{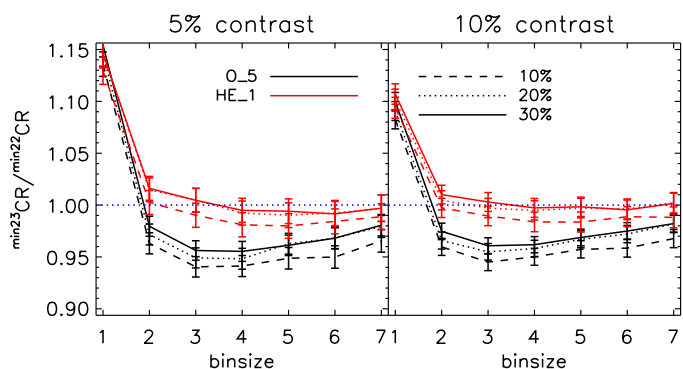}}
  \caption{Ratio between CR at the two solar minima for different binsizes. The blue horizontal dotted line indicates the unity line. The dashed, dotted and solid lines are corresponding to the three versions of data exclusion, corresponding to 10\%, 20\% and 30\% data exclusion due to existence of bright features inside the field of view, according to progressively stricter criteria. The error bars represent the 1$\sigma$ error in the ratios.}
  \label{cnt_bp_allbins}
\end{figure}
\begin{figure*}
\centering
{\includegraphics[width=18cm]{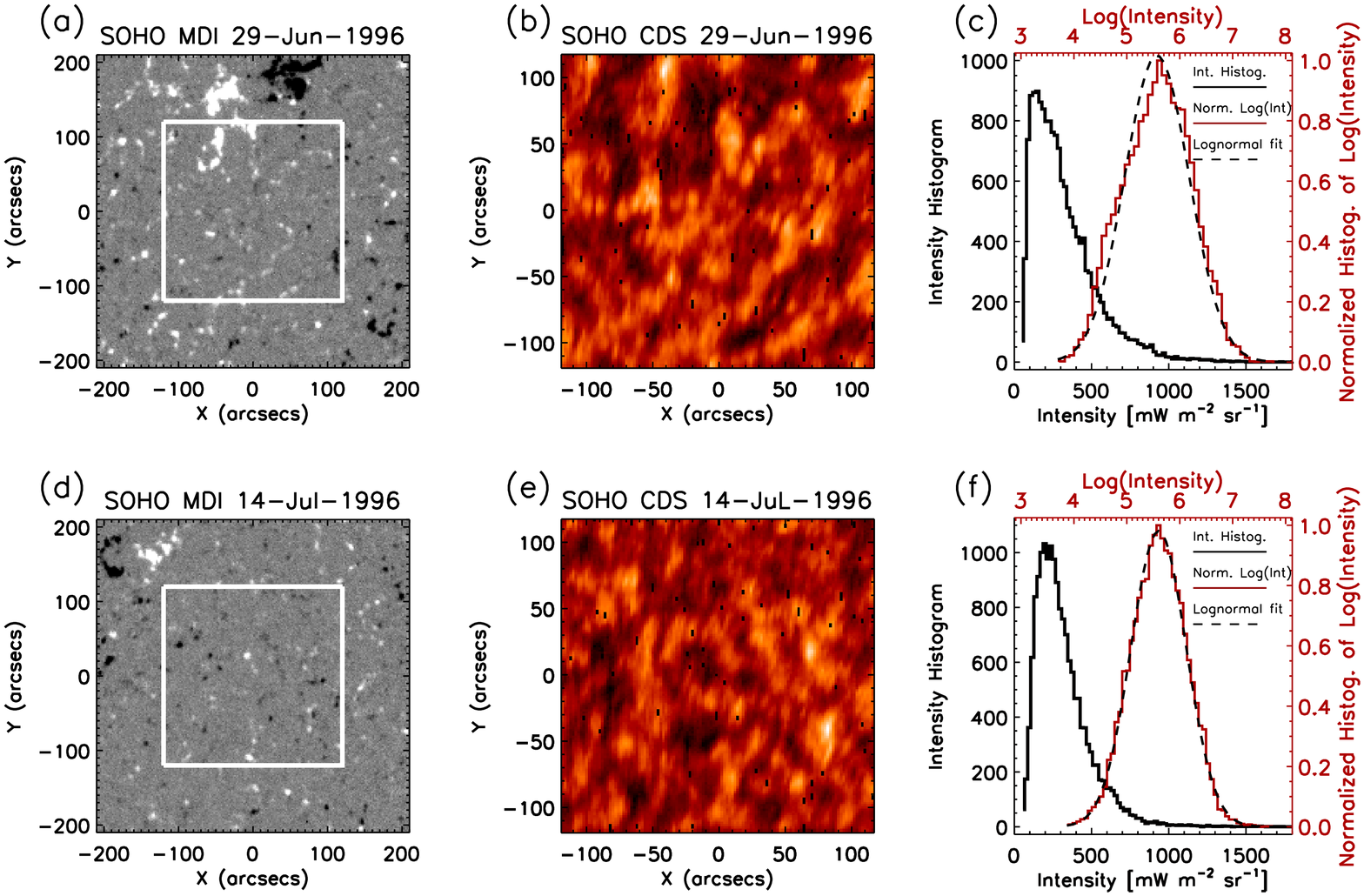}}
  \caption{From left to right:  MDI magnetograms, \ion{O}{V} integrated intensity scans and intensity distribution functions obtained from the intensity images. The solid black lines in panels c and f show the intensity histograms on a linear intensity scale scale (corresponding to the axes on bottom and left side of panels c and f). The solid red lines are the normalized histograms of the logarithm of the intensity (corresponding to the axes on top and right side of panels c and f). The dashed lines are lognormal fits to the normalized histograms. The surface area visible in CDS images is marked by the white rectangles in the MDI magnetograms.}
  \label{lognormal_sample}
\end{figure*}
To have a direct comparison between the two minima, we computed the ratio of the mean CR of minimum 23 to that of minimum 22 ($^{min23}\overline{CR}$\:/\:$^{min22}\overline{CR}$) for both spectral lines. This ratio of CRs is illustrated in Fig.~\ref{cnt_bp_allbins} as a function of binsize. The ratios are either close to or below the unity-line for all binsizes except binsize 1. This is related to the fact that the darker pixels are dominated by noise for the data taken during solar minimum 23 due to the drop in the instrument responsivity (see Fig.~\ref{adc}). Therefore, when making Gaussian fits to individual spectra in darker areas, the signal to noise ratio is so low that often what is being fitted is noise rather than the spectrum. Therefore, the line intensities in these areas tend to be underestimated. This effect is larger for the data of minimum 23 and explains why the CRs are larger during solar minimum 23. 
In addition to binsize values, the plots in the figure also highlight the effect of data exclusion. Three different versions of data exclusion were applied to examine how sensitive the ratios are to the data exclusion. In version 1, almost 10\% of the images containing the most obvious bright structures (chosen visually) were excluded from the data set. Versions 2 and 3 correspond to 20\% and 30\% data exclusion, respectively. They represent progressively stricter rejection criteria. 
It is evident from the plots that by excluding 20\% of the data, the \ion{He}{I} has already reached a stable state since the ratios for versions 2 and 3 lines almost overlap and excluding more data points does not change the minimum-to-minimum ratios for any binsize. This is different for \ion{O}{V}, where excluding more data keeps pushing the ratios toward the unity line. 

The plots also show that the minimum-to-minimum ratios are less than 1 for all binsizes (except binsize 1) in case of \ion{O}{V} and are very close to unity in \ion{He}{I}. This suggests that the intensity distribution of \ion{O}{V} underwent significant changes between the last two solar minima while that of the upper chromospheric \ion{He}{I} did not.

\subsection{Intensity distribution of the quiet Sun radiation and lognormal fitting}\label{lognormal}
Various studies have shown that the intensity distribution of the quiet Sun in the chromosphere and transition region has a skewed shape, with a pronounced peak at relatively low intensities and an extended tail at higher intensities. This skewed distribution has been modelled mainly by either a sum of two Gaussian functions (\citeads{1976SoPh...46...53R}, \citeads{1998A&A...335..733G}) or a single Lognormal function (\citeads{1999ApJ...512..992G}, \citeads{2000A&A...362..737P}, \citeads{2007A&A...468..695F}), each representing a different hypothesis about the source of the radiation in quiet Sun regions. These two models imply different heating scenarios. A two-component function assumes two intrinsically different heating mechanisms for radiation originating in the solar network and internetwork, while a single-component model assumes similar heating mechanisms for both network and internetwork (\citeads{2000A&A...362..737P}). The common agreement today is that the quiet Sun radiation follows in general a lognormal distribution function. Thus \citetads{2000A&A...362..737P} showed that despite its fewer free parameters, a lognormal reproduced the observed distributions significantly better than the sum of two Gaussians. 

It is believed that a system variable can be modelled with a lognormal distribution if it is a multiplicative product of many independent forces acting on that system (\citealt{Gronholm2007659}, \citealt{LIMPERT2001}).  These expressions can be applied to solar physics through, e.g., the nanoflare heating model, which is a potential mechanism for coronal heating (\citeads{1983ApJ...264..642P}, \citeyearads{1988ApJ...330..474P}). The consistency between lognormal distribution of transition region radiation at the quiet Sun and nanoflare heating of the corona has been studied by \citetads{2007A&A...462..311P} and \citetads{2008A&A...492L..13B}.

As mentioned earlier, in order to study the variation of the radiance distribution over time, we fit the individual distributions with a lognormal function, extract the best fit parameters and investigate their long-term variability. The lognormal distribution function can be mathematically defined by replacing the independent variable (here intensity \textit{I}) in the normal distribution function with its logarithm. In other words, if $\log{(I)}$ is normally distributed, the variable \textit{I} is lognormally distributed: 

\begin{equation}
\label{logn}
P(I) = \frac{N\textsubscript{0}}{{\sigma I\sqrt {2\pi } }}      \exp{\left(-\frac{     {  \left( {\ln(I) - \mu } \right)^2 }    }{    2\sigma ^2 }\right).}
\end{equation}

The above equation shows a lognormal probability density $P(I)$ as a function of intensity. Here $\mu = \langle \ln{(I)}\rangle$, $\sigma = \sqrt{\mathrm{Var}{(\ln{(I))}}}$ and $N\textsubscript{0}$ is the normalization factor. The $\mu$ and $\sigma$ are the mean and standard deviation of $\ln{(I)}$. We use natural logarithm in the definition of the lognormal function. Using another base to the logarithm only rescales the fitting parameters, but eventually returns the same family of distributions. The back-transformed parameters are then defined as 

\begin{equation}
\label{param}
\mu^* = e^{\mu} \;,\; \sigma^* = e^{\sigma}
\end{equation}

and called scale parameter and multiplicative standard deviation or shape parameter, respectively (\citealt{LIMPERT2001}). 

Fig.~\ref{lognormal_sample} illustrates two intensity images in the \ion{O}{V} line, their corresponding histograms and two MDI magnetograms of a slightly larger solar area, encompassing the area scanned in \ion{O}{V}. The upper row corresponds to a date on which part of an active region passed through the field of view, while the bottom row depicts quiet Sun conditions. Each plot consists of two histograms: the actual intensity histogram shown by the solid black line and the normalized histogram of the natural logarithm of the intensity (\textit{$\log(I)$}) represented by the solid grey line. The dashed lines are Gaussian fits to the normalized histogram of the logarithm of the intensity distribution. A comparison between the two plots reveals how significantly the distributions can deviate from a lognormal distribution when a particularly bright structure, such as a part of an AR, appears in the field of view. $\chi^2$ values, as will be explained later, were applied to exclude the non-quiet Sun data from the analyses as an alternative approach to visual inspection.

\begin{figure}
    \resizebox{\hsize}{!}{\includegraphics{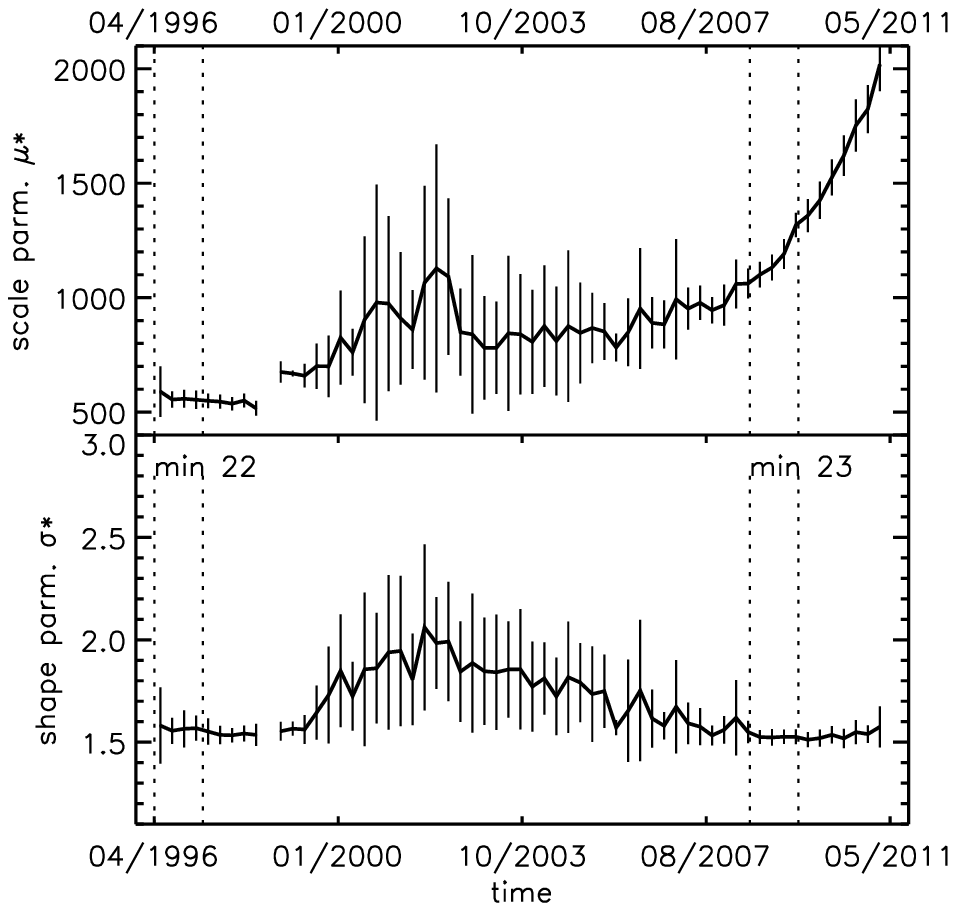}}
  \caption{Long-term variations in the scale (upper panel) and shape (lower panel) parameters of the lognormal fit to the intensity distribution of the \ion{He}{I} spectral line. The solid lines and vertical error bars correspond to 90-day average values and standard deviations, respectively.}
  \label{scl_shp_heI}
\end{figure}

The scale and the shape parameters of the individual intensity distributions were calculated according to the $\mu$ and $\sigma$ values of the normal fits, as formulated before. Note that the scale parameter is sensitive to the instrumental corrections. Referring to the mathematical definitions of the shape and scale parameters given above, multiplying an intensity image to a given number (as a correction factor) shifts the position of the peak of the histogram and consequently changes the scale parameter, but not the shape parameter. This can be seen in Fig.~\ref{scl_shp_heI}, which exhibits the variation in scale and shape parameters of the intensity distributions of the \ion{He}{I} spectral line. The scale parameter shows a similar trend to that of the average intensities of \ion{He}{I} shown in Fig.~\ref{avInt}, while the shape parameter does not depict any systematic trend. 

The plot of the shape parameter in Fig.~\ref{scl_shp_heI} also demonstrates an increase as we approach the solar maximum. This happens since the overall emergence of active areas during the solar maximum gives rise to the number of bright pixels in the tails of the radiance distributions. \citeads{2000A&A...354L..71S} found similar increasing trends in the radiance of various ultraviolet lines between 1996 and 2000. They showed that the network and internetwork radiances increase uniformly over the solar cycle. However, whether such an increase occurs indeed uniformly over network and internetwork can only be seen in daily average intensities, such as those in Fig.~\ref{avInt} (from the minimum to the maximum of cycle 22). However, we can not verify this finding due to the uncertainties in the absolute intensity values.

\begin{figure*} 
\sidecaption
\centering
  \includegraphics[width=12cm]{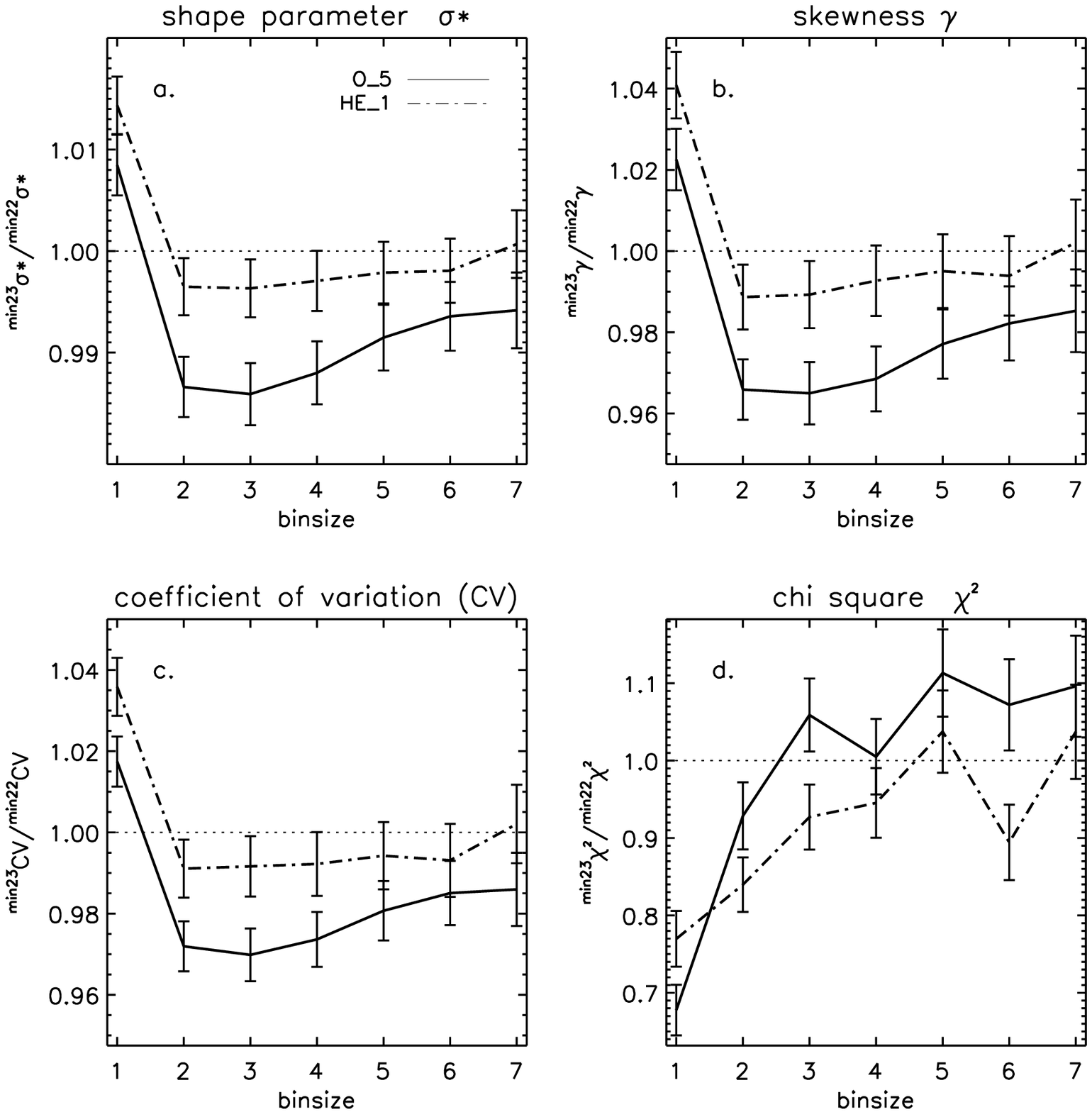}
  \caption{Cycle 23 activity minimum to cycle 22 activity minimum ratios of various statistical parameters of the lognormal fits. The solid and dash-dotted lines represent the ratios in \ion{O}{V} and \ion{He}{I} spectral lines, respectively. The error bars represent the 1$\sigma$ error in the ratios.}
  \label{lognormal_parm}
\end{figure*}

Various statistical parameters can be defined using the $\mu$ and the $\sigma$ values. We do not analyse the parameters that are dependent on $\mu$  (e.g., scale parameter, mean and mode) for the same reason as we don't analyse the average intensities. These parameters can be used to improve the instrument calibration, as was done by \citetads{2010A&A...518A..49D}. We decided to follow a different approach here. Parameters that only depend on $\sigma$ are best for studying long-term variations. Four such statistical parameters were chosen for further study, the shape parameter $\sigma^*$, the skewness $\gamma$, the coefficient of variation \textit{(CV)} and the $\chi^2$ value of the fits. The CV parameter is a measure of the variable dispersion, while the skewness is a measure of the asymmetry of the distribution. In general, the distribution is more asymmetric and dispersed for larger $\sigma$, while a smaller $\sigma$ implies that the distribution is more symmetric. Skewness and CV are computed as follows:

\begin{equation}
\gamma = \left(e^{\sigma^2}+ 2\right) \sqrt{e^{\sigma^2}-1} \quad \text{(Skewness)}
\end{equation}
\begin{equation}
CV = \sqrt{e^{\sigma^2}-1}  \qquad \qquad  \text{(Coefficient of Variation)}
\end{equation}

Fig.~\ref{lognormal_parm} displays the ratios of the minimum of cycle 23 values to those during cycle 22 minimum of various parameters for different binsizes. Plotted are such ratios of $\sigma^*$, $\gamma$ and \textit{CV} averaged over the periods of time defined as minima. Excluding the $\chi^2$ plot which gives a measure of the goodness of the fits, all other plots depict lower than unity ratios for all binsizes greater than 1, specially for \ion{O}{V}. This suggests that the intensities follows a less asymmetric distribution during solar minimum 23, i.e., the intensity distributions are closer to a normal distribution. This effect is less pronounced in \ion{He}{I}, with closer-to-unity ratios compared to the \ion{O}{V} spectral line. This implies that the mean values at the two ends of the intensity distributions are closer in magnitude, so that this result is in agreement with the 5\% and 10\% CRs being smaller at minimum 23, as shown in Fig.~\ref{cnt_bp_allbins}. Here again the exception is the binsize 1, which shows a similar behaviour to that of the CRs. The $\chi^2$ plot helps to determine which amount of binning is most appropriate for comparing the two minima (where the $\chi^2$ is the same in both minima). The best consistency between the data for two minima, in terms of signal to noise ratio, can be achieved for binsizes 3 and 4 for both spectral lines.

\begin{figure*} 
\sidecaption
\centering
  \includegraphics[width=12cm]{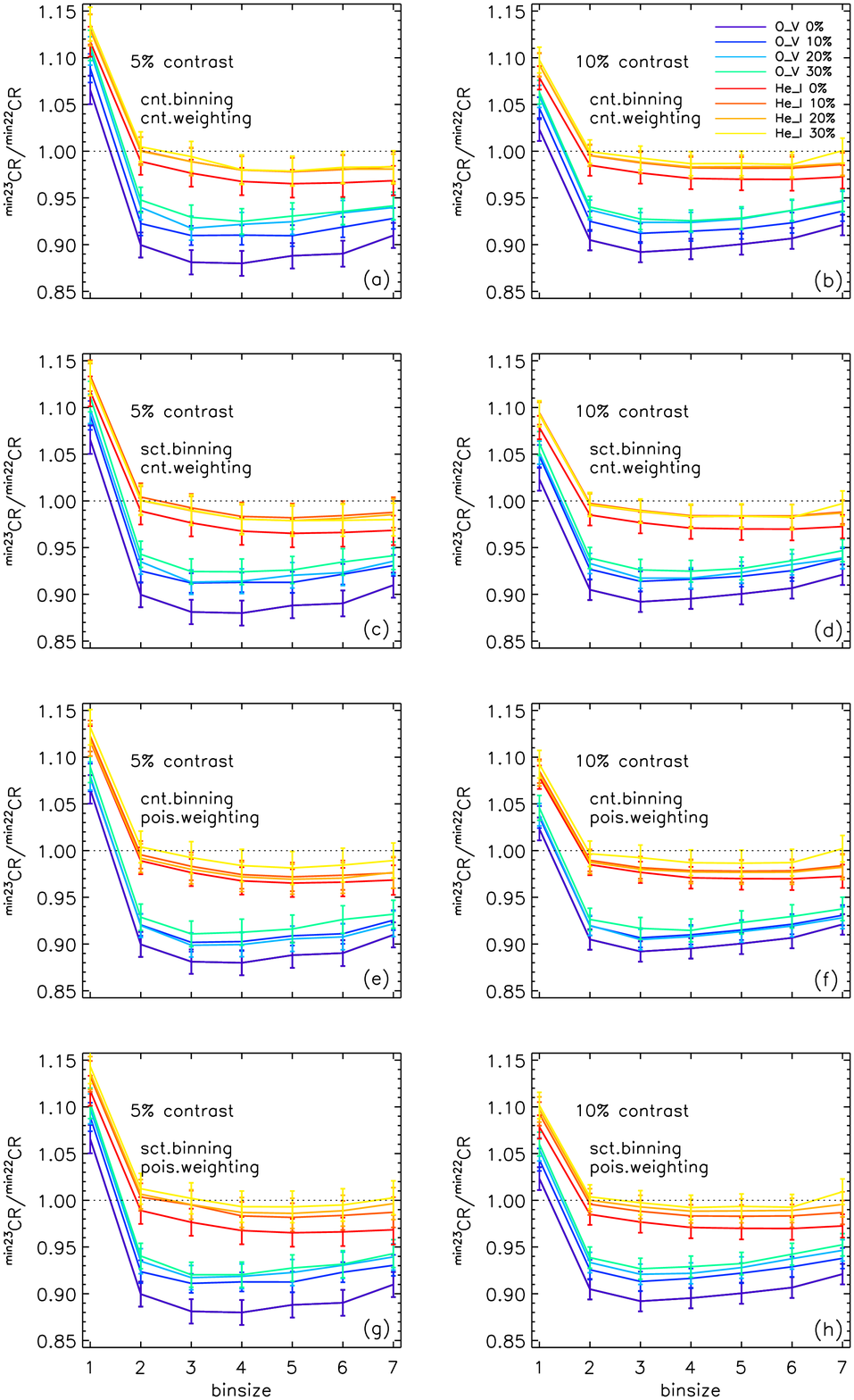}
  \caption{Minimum-to-minimum ratio of the 5\% and 10\% CRs. Each panel, consists of the ratios for both spectral lines and for different amounts of data exclusion, based on the $\chi^2$ values of a lognormal fit to radiance distributions. The effect of two different methods of determining the histogram binsize and weighting them for fitting has been emphasized. The abbreviations \textit{cnt} and \textit{sct} and \textit{pois} are for constant, Scott's rule and Poisson, respectively. The differently coloured lines are labelled in panel b.}
  \label{CONTRASTS_everything3}
\end{figure*}

As already shown in Fig.~\ref{lognormal_sample}, the intensity distributions of the images with any trace of activity exceeding that of the typical quiet Sun deviate from a lognormal. The $\chi^2$ values of the lognormal fits, as mentioned before, can therefore also be utilized as a potential reference for excluding scans that do not cover purely quiet Sun. Fig.~\ref{CONTRASTS_everything3} shows the minimum-to-minimum ratios of the 5\% and 10\% CRs, similar to the one which is already shown in Fig.~\ref{cnt_bp_allbins}, but this time instead of excluding the data by visual inspection, we selected them according to the $\chi^2$ of their lognormal fits. Here again three different versions were introduced corresponding to the number of data points which were excluded. 0,10, 20 and 30\% in the images label implies that this fraction of the intensity images with highest $\chi^2$ values, was removed before the CRs were computed. In order to check if the ratios are sensitive to the histogram bin width and/or to the weighting of the bins for lognormal fitting, we tested two different methods for defining the histogram bin width and two methods to weight the histogram bins. The bin widths were either given a fixed value equal to 20 mW/m\textsuperscript{2}/sr for the entire data set and for both spectral lines or alternatively determined according to Scott's normal reference rule for each individual histogram. Scott's normal reference rule estimates the optimal bin width for a histogram by assuming the underlying distribution to be a Gaussian and by minimising the integrated mean square error (IMSE) of the histogram \citep{SCOTT01121979}. For lognormal fitting, the bins were either given a constant weight or a statistical (Poisson) weight. The four different combinations of these methods for the two CRs are illustrated in Fig.~\ref{CONTRASTS_everything3}. In most cases, there is a remarkable gap between the 0\% line and the other lines, whereby '0\% line' refers to the ratio of CRs obtained without removing any data (i.e. keeping also the brightest scans, most strongly affected by active regions). The fact that the 10, 20 and 30\% lines almost overlap each other, implies that the ratio between the CRs approaches a saturation level when going to larger exclusion percentages. This was less evident for \ion{O}{V} in Fig.~\ref{cnt_bp_allbins}, where the exclusion was done by visual inspection. Here again, regardless of which method for determining the bin widths and bin weights is used, the \ion{He}{I} line does not show any significant difference between the two minima. The \ion{O}{V} CRs, on the other hand, exhibit a drop of approximately 8\% during minimum 23 compare to that of minimum 22, which is almost twice the change we saw in the CRs when excluding the bright days by visual inspection.

\section{Discussion and conclusion}
The degradation of the CDS instrument's photometric sensitivity makes the absolute intensity of spectral lines unreliable for long-term investigations of subtle solar changes such as the difference in brightness between the minima following cycles 22 and 23. A qualitative study of the parameters independent of absolute intensity, is a possible alternative that we consider here. For this purpose, we defined two contrast ratios (CRs), as the ratio between the mean intensity of the brightest and darkest parts of the intensity images (see Fig.~\ref{crh}) . This way we cancelled out the effect of correction and calibration factors. We also fitted lognormal distribution functions to the intensity distributions and extracted the fit parameters. Eventually a comparative study of all these parameters over the last two solar minima was carried out to investigate the variations in the radiance distribution. We also excluded the images with bright structures in the field of view to concentrate only on the quiet Sun area. The exclusion procedure was based either on visual inspection (Fig.~\ref{4BP}) or on the goodness of the lognormal fits to the intensity histograms (Fig.~\ref{lognormal_sample}). In addition, the NIS raster scans were down-sampled (binned) prior to any further analyses to compensate for the effect of reduced sensitivity of the instrument by increasing the signal to noise ratio over darker areas. This does not effectively reduce the instrument spatial resolution up to binsize 3, as the instrument PSF is almost three times larger than the CDS actual pixel size (see section~\ref{psf}). Therefore, binsizes 3 and 4 are believed to be the optimal values for down-sampling in order to have maximum signal to noise ratio and minimum loss of spatial resolution. This is in agreement with previous work by \citetads{2000A&A...353.1083B}.

The CRs in \ion{O}{V} exhibit a remarkable drop over solar minimum 23. The magnitude of the reduction, however, is sensitive to the data exclusion methods. From Fig.~\ref{cnt_bp_allbins}, when considering binsizes 3 and 4, exclusion of 30\% of the images by visual inspection reveals a drop of around 4\% in the minimum-to-minimum trend. The drop is almost doubled when the exclusion is based on the $\chi^2$ of the lognormal fits, displayed in Fig.~\ref{CONTRASTS_everything3}. The latter figure shows that no substantial changes occur when rejection levels larger than 10\% are used.

The CRs, are in principle a measure of contrast between the network and internetwork radiances. The increase of the ratio of the CRs as the number of excluded bright images is increased, is due to the fact that we are left with progressively "purer" quiet Sun when more and more images with bright features are excluded. The \ion{O}{V} radiance distributions undergo a shift from a skewed form to a more symmetric shape during the very quiet minimum of cycle 23. Such a variation of the distribution manifests itself in terms of a decrease in the shape and skewness parameters illustrated in Fig.~\ref{lognormal_parm}, where the skewness of the intensity distribution of the \ion{O}{V} line shows a drop of 3.5\% over solar activity minimum 23. The smaller dispersion in \ion{O}{V} intensity distribution, as quantified by the coefficient of variation (panel \textit{c} of Fig.~\ref{lognormal_parm}), during activity minimum 23 suggests that either the thermal structure of the internetwork has changed between the minima following cycles 22 and 23, or the area covered by the network or the brightness of the network. The latter two possible explanations relate to the amount of magnetic flux. Lower magnetic flux is also found to be the cause of the lower TSI (we refer to TSI composites PMOD and ACRIM, although the RMIB and TIM composites do not show a decrease) during the minimum of cycle 23 (cf. \citeads{2012A&A...541A..27B}; \citeads{2014A&A...570A..85Y}). On the other hand, the studies carried out by \citetads{2011CoSka..41..113F} and \citetads{2010ApJ...711L..84T}, show also a drop in the 10.7\,cm flux up to 4\% over minimum 23 with respect to the previous one. We note that the 10.7\,cm radio flux is a global solar activity factor and it is not trivial to compare it directly with our results, which are entirely based on local observations of the solar disc centre. 

The higher CRs in transition region \ion{O}{V} line during solar minimum 22 can also be attributed to the presence of active network elements, being dispersed within the quiet Sun. The active network results from the decay of active regions (\citeads{1984ApJ...282..776S, 1998ApJ...496..998W, 1999ApJ...511..965W}) and may last for several solar rotations before diffusing into the quiet Sun (\citeads{2000JGR...10527195W}). The exclusion criteria for rejecting images with bright structures may have caused us to miss the active network contribution in the quiet Sun radiance distribution.

The \ion{He}{I} spectral line, in contrast to the \ion{O}{V} line, shows almost no minimum-to-minimum variations. This is evident when referring to the CRs in Figs. \ref{CONTRASTS_everything3} and \ref{cnt_bp_allbins}. Both 5\% and 10\% CRs at binsizes 3 and 4 are very similar in the last two solar minima and their minimum-to-minimum ratios are within 1\% of the unity line. It is also interesting to note that the \ion{He}{I} line is less sensitive to the exclusion of data due to the presence of magnetic activity. Excluding only 10\% of the images with the most obvious bright structures, is sufficient for the total brightness not to be affected by the remaining traces of activity. This is because the contrast between network and internetwork is larger in the transition region \ion{O}{V} line than in the chromespheric \ion{He}{I} line. In addition, \citetads{2000JGR...10527195W} show that the contrast for the active network is more enhanced for transition region emission (such as in \ion{He}{II} 304\,\AA) than it is for chromospheric or coronal emission. This is also the likely reason why the \ion{O}{V} intensity distribution has a more skewed shape over minimum 22.

These findings are compatible with those of \citetads{2010MmSAI..81..643L}, who found that the chromospheric Ca II K line, measured over solar disc centre, is constant (within 1\%) between the last three solar activity minima. This is important, for the Ca II index is usually regarded to display a relatively tight relationship with the photospheric magnetic field (\citeads{1975ApJ...200..747S}; \citeads{1989ApJ...337..964S}; \citeads{2009EGUGA..1111903L}). The \ion{He}{I} and Ca II K results imply a certain level of decoupling between the chromospheric radiation and the photospheric field at least at very low level of activity.

In a recent study by \citetads{2014arXiv1401.7570A}, the authors also analysed the radiance distribution over solar cycle 23 by investigating the full-disc and central meridian observations of the CDS studies USUN and SYNOP, respectively. They concluded that the quiet Sun radiance has undergone no significant changes during solar cycle 23, regardless of which spectral line is considered. Although we came to the same conclusion for \ion{He}{I}, we found a minimum-to-minimum variation in the radiance distribution of the \ion{O}{V}. Nevertheless, the values they found as the average widths of the lognormal fits over the full cycle are in good agreement with our findings (when rescaled from natural logarithm to normal logarithm). We measured average width values for histograms in \ion{He}{I} and \ion{O}{V} to be equal to $0.18\pm0.01$ and $0.22\pm0.02$ dex, respectively. The values reported in the study of \citetads{2014arXiv1401.7570A} are $0.16\pm0.015$ dex for \ion{He}{I} and $0.19\pm0.02$ dex for \ion{O}{V}. The small difference might be due to the specific fitting approach applied by \citepads{2014arXiv1401.7570A}, in which only the peak of the histograms, corresponding to the internetwork area, are fitted and the contribution of the brighter network is ignored. 

In the model introduced by \citeads{1986SoPh..105...35D}, a two-component mechanism is assumed to be responsible for producing the emission in the transition region (TR): a downflow of the heat within the magnetic funnels which expand upwards throughout the network structures connecting to corona, and the internal heating of the gas inside low-lying small loops, which are magnetically insulated from the corona. Each of these two components connects the radiance formation in the TR to the chromosphere and corona. 

The difference in behaviour of the chromospheric and TR lines found here may have different origins. One possibility is that the \ion{He}{I} line reacts less strongly to the presence of a magnetic field than the \ion{O}{V} line. Another is that the heating of the quiet chromosphere is much more strongly determined by acoustic waves than the TR. As a result, in \ion{He}{I} we are basically seeing basal flux in the very quiet Sun, while in TR lines a more magnetically dominated heating mechanism is active. The former does not change from minimum to minimum, while the latter does.

\begin{acknowledgements} 
SOHO is a project of international collaboration between ESA and NASA. We would like to thank the CDS team for developing and operating the instrument, as well as SOHO/MDI and SOHO/EIT consortia for their data. We thank all the people who provided the CDS manual scripts and software notes. The authors also thank T. Woods for helpful suggestions. In addition, we used SolarSoft library for the majority of the analyses. F. Shakeri would like to acknowledge the International Max-Planck Research School for Solar System Science run jointly by the
Max Planck Society and the Universities of G\"ottingen, for supporting this project. This work was partly supported by the BK21 plus program through the NRF founded by the Ministry of Education of Korea. 
\end{acknowledgements}

\bibliographystyle{aa}    
\bibliography{my_bib1.bib}    

\end{document}